%% file: main.tex
\documentclass[10pt,twocolumn,letterpaper]{article}

\usepackage[pagenumbers]{cvpr} 

\input{preamble}

\definecolor{cvprblue}{rgb}{0.21,0.49,0.74}
\usepackage[pagebackref,breaklinks,colorlinks,citecolor=cvprblue]{hyperref}

\usepackage{array,multirow}
\usepackage{kotex}
\usepackage{mathtools}
\usepackage{arydshln}
\usepackage{amssymb}
\usepackage{bbding}
\usepackage{pifont}
\usepackage{algorithmic}
\usepackage{algorithm}
\usepackage{bm}

\usepackage{stfloats}
\fnbelowfloat
\usepackage{colortbl}
\usepackage{adjustbox}
\usepackage{colortbl}

\newcommand\blfootnote[1]{%
  \begingroup
  \renewcommand\thefootnote{}\footnote{#1}%
  \addtocounter{footnote}{-1}%
  \endgroup
}
\usepackage[accsupp]{axessibility} 
\usepackage{appendix}

\def\paperID{4389} 
\def\confName{CVPR}
\def\confYear{2024}

\title{Data-Efficient Unsupervised Interpolation \\ Without Any Intermediate Frame for 4D Medical Images}

\author{
JungEun Kim\textsuperscript{1*} \quad Hangyul Yoon\textsuperscript{1*} \quad  Geondo Park\textsuperscript{1} \quad Kyungsu Kim\textsuperscript{2\dag} \quad Eunho Yang\textsuperscript{1,3}\vspace{0.05in}\\
\textsuperscript{1}Korea Advanced Institute of Science and Technology (KAIST)\vspace{0.02in}\\
\textsuperscript{2}Massachusetts General Hospital and Harvard Medical School \quad \textsuperscript{3}AITRICS\\
{\tt\small\{jungeun122333, hangyulmd, geondopark, eunhoy\}@kaist.ac.kr \quad kskim.doc@gmail.com}}

\begin{document}
\maketitle

\begin{abstract}
4D medical images, which represent 3D images with temporal information, are crucial in clinical practice for capturing dynamic changes and monitoring long-term disease progression. However, acquiring 4D medical images poses challenges due to factors such as radiation exposure and imaging duration, necessitating a balance between achieving high temporal resolution and minimizing adverse effects. Given these circumstances, not only is data acquisition challenging, but increasing the frame rate for each dataset also proves difficult. To address this challenge, this paper proposes a simple yet effective \textbf{U}nsupervised \textbf{V}olumetric \textbf{I}nterpolation framework, UVI-Net. This framework facilitates temporal interpolation without the need for any intermediate frames, distinguishing it from the majority of other existing unsupervised methods. Experiments on benchmark datasets demonstrate significant improvements across diverse evaluation metrics compared to unsupervised and supervised baselines. Remarkably, our approach achieves this superior performance even when trained with a dataset as small as one, highlighting its exceptional robustness and efficiency in scenarios with sparse supervision. This positions UVI-Net as a compelling alternative for 4D medical imaging, particularly in settings where data availability is limited. The code is available at \href{https://github.com/jungeun122333/UVI-Net}{UVI-Net}.
\end{abstract}

\vspace{-0.2cm}
\section{Introduction}
\label{sec:intro}
\blfootnote{\textsuperscript{*}Eqaul Contribution \; \textsuperscript{\dag}Correspondence to}Video Frame Interpolation (VFI) has been a cornerstone in the realm of video processing, enriching motion visualization by generating intermediate frames. This method primarily relies on intermediate frame supervision, where known frames are used as references to create new intermediate frames. However, applying these VFI methods to 4D medical imaging is not trivial. While the principles of frame interpolation hold the potential for enhancing medical diagnostics and treatments~\cite{xi2007defining, bellec2020itv, wang2009dosimetric, jeung2012myocardial, mcveigh2001imaging, hor2011magnetic, mcveigh2018regional}, the unique constraints and requirements of medical imaging present challenges. 

One significant challenge lies in obtaining a sufficient dataset. Unlike general domain videos, 4D medical images are captured for specific clinical purposes from a relatively small pool of individuals. Similarly, acquiring intermediate frames per image is also hampered by limitations and risks associated with medical imaging modalities. 

Computed tomography (CT) exposes patients to elevated radiation levels, potentially increasing the risk of secondary cancer~\cite{wang2023risks}. Similarly, magnetic resonance imaging (MRI) faces the obstacle of lengthy scan times, lasting up to an hour~\cite{sartoretti2019reduction}, presenting both logistical challenges and issues related to patient comfort. Furthermore, the quality of ground truth intermediate frames in medical imaging is often compromised due to factors such as patient movement, unstable breathing, and the difficulty of maintaining a stable position during prolonged scans \cite{caines20224dct, mizuno2021preoperative}, limiting data variety and accessibility for research.

In light of these challenges, we present the following question: ``Can a VFI model be trained without depending on \emph{any} ground truth intermediate frames?". Unlike other previous unsupervised approaches in the 2D general domain~\cite{reda2019unsupervised, seung2020unsupervised, liu2019deep} that interpolate frames given the multiple frame sequences, we address the task of freely interpolating between two given frames \textit{without any intermediate frames.} To achieve this, we propose a straightforward yet effective framework to VFI in medical imaging. By interpolating the flow between two frames with a two-stage process and cycle-consistency constraint, our framework can effectively operate even with a video composited with two frames (i.e., only images of the start and end points exist), entirely in an unsupervised manner. In the initial stage, virtual samples are generated from the two real input images. Subsequently, the real images are reconstructed based on these virtual intermediate samples. This reconstruction process incorporates the candidate images and warped contextual information in multiple scales from the input images. Through this cyclic interpolation approach, we successfully minimize discrepancies between the generated and the actual images by using the real images as a form of pseudo-supervision.

Our proposed method has achieved state-of-the-art results in unsupervised VFI for 4D medical imaging, outperforming the existing techniques with a substantial gap. Our approach also consistently outperforms even for existing supervised methods. Remarkably, our model shows competitive or even superior performance when trained with a minimal training dataset size of just one, contrasting with other baselines that require full datasets, typically exceeding 60 in size. Additionally, the unsupervised nature of our model allows for further performance enhancements through instance-specific optimization. This process involves briefly fine-tuning the model using each test sample during the inference stage, potentially yielding even more refined results.

In summary, our contributions are three-fold:
\begin{itemize}
    \item We introduce a simple yet effective unsupervised VFI approach for 4D medical imaging. Our methodology leverages cycle consistency constraints within the temporal dimension, thereby obviating the need for ground truth data typically required for interpolated images. 
    
    \item Our approach achieves state-of-the-art performance, surpassing other unsupervised and supervised interpolation methods. This is accomplished without the instance-specific optimization, which could be employed as a viable option to enhance performance.
    
    \item The robustness of our model is particularly evident under conditions of limited data availability, as demonstrated by the increasing performance margin relative to other methods when the dataset size is reduced.
\end{itemize}
\vspace{-0.1cm}

\section{Related Works}
\label{sec:related work}
\subsection{Video interpolation}
\label{subsec:related work video interpolation}
Many studies in the field of video interpolation have been conducted, with a significant emphasis on frame rate upsampling for natural scene videos~\cite{meyer2018phasenet, peleg2019net, niklaus2017video}. These studies typically rely on ground truth intermediate frames for training~\cite{jiang2018super, xiang2020zooming, zhang2023extracting, park2023biformer, jin2023unified, zhou2023exploring, niklaus2020softmax, chen2022videoinr}. While some studies have explored alternative approaches that do not rely on ground truth intermediate frames, they involve synthesizing frames between a given sequence of intermediate frames~\cite{reda2019unsupervised, seung2020unsupervised, liu2019deep} or utilize the information from specialized devices, such as event camera~\cite{he2022timereplayer}. Consequently, applying these methods in settings like our study presents a challenge, as there are no intermediate frames available for synthesis. Furthermore, validation of these methods is restricted to 2D frames, and they encounter challenges when directly applied to volume sequences. This is primarily attributable to the markedly lower availability of intermediate frames within such datasets, as elucidated in \cref{tab:dataset comparison}. 
\input{table/compare_dataset} 
\vspace{-0.4cm}

\paragraph{Medical 4D image interpolation.}
To address the above challenges, frame interpolation methods specifically focused on 4D medical images are driven. Several recent works~\cite{guo2020spatiotemporal, guo2021unsupervised} have attempted to interpolate medical 4D images, but these methods rely on the availability of ground-truth intermediate images for training. Although~\citet{kim2022diffusion} proposed an interpolation approach without using the authentic intermediate frames, they do not incorporate an unsupervised learning technique for the interpolated samples. Instead, their method involves a post-hoc multiplication of the flow calculation model, which is prone to spatial distortion. This weakness arises since the underlying network does not account for the structural smoothness between two samples during network training. As a result, the scaled calculated flow fails to capture the spatial continuity of intermediate samples beyond the samples provided by authentic frames. Furthermore, since the method focus solely on warping without incorporating image synthesis, they encounter specific issues if a voxel is displaced to a new location without replacement at the original site. Specifically, it results in the voxel appearing twice in the backward-warped frame~\cite{lee2022enhanced, lu2020devon}, or a hole at the original location in the forward-warped frame~\cite{niklaus2018context}. To overcome the limitation of nonexistent training for intermediate images and warping procedure, we propose a novel network incorporating pseudo-supervision, including an image synthesis network to ensure the integrity of intermediate images.

\subsection{Learning optical flow}
Optical flow learning is crucial in the video and medical domain. Various learning methods have been extensively investigated~\cite{yang2017quicksilver, sokooti2017nonrigid, cao2018deep} aiming to estimate optical flow. However, they require a ground truth optical flow for training, which is limited in availability. To address this limitation, some methods ~\cite{beg2005computing, balakrishnan2018unsupervised, voxelmorph2019, lei20204d, kim2022diffusemorph, kuang2019cycle, kim2021cyclemorph, karani2019image, joshi2023r2net, jia2023fourier, wolterink2022implicit} have been developed to compute the similarity between the warped image and a fixed reference to train networks, allowing training without ground truth optical flow.

\section{Background}
\label{sec:background}
We first briefly introduce the necessary background on the flow calculation model in~\cref{subsec:flow calculation model} and the existing unsupervised interpolation approaches in~\cref{subsec:baseline approach}. 

\subsection{Flow calculation model}
\label{subsec:flow calculation model}
Suppose we are given two input images \(I_0\) and \(I_1\) at time $T=0$ and $T=1$, respectively. Our main objective is to predict the intermediate image \(\hat I_t\) at time \(T=t\) within the range of 0 to 1, given \(I_0\) and \(I_1\), without explicit supervision. An intuitive approach is to train a neural network to directly generate voxel values of \(\hat I_t\) without explicitly computing coordinate transformation. However, the generation models such as generative adversarial networks (GAN)~\cite{dai2020multimodal, nie2017medical, nie2018medical, yang2018unpaired} typically require a large amount of training data, making them impractical for the medical domain where data is limited. In contrast, flow calculation models~\cite{jiang2018super, xiang2020zooming, reda2019unsupervised} can generate 3D images using only two real input images. Given these advantages, we employ flow-based methods for this task, as they are capable of generating 3D images using only two real input images.

Flow-based interpolation approaches employ a flow calculation model \(\mathcal{F}^\theta\) with model parameters \(\theta\) to obtain a coordinate transformation map between two target samples. Given \(I_0\) and \(I_1\), the flow calculation model \(\mathcal{F}^\theta\) takes \(I_0\) and \(I_1\) as sequential inputs and provides a coordinate transformation map \(\phi^\theta_{0 \rightarrow 1}\). The objective of the flow calculation model \(\mathcal{F}^\theta\) is to warp \(I_0\) into \(\hat I_{0 \rightarrow 1} := I_0 \circ \phi^\theta_{0 \rightarrow 1}\) such that it matches \(I_1\), where \(\circ\) indicates spatial transformation. 

To train flow calculation models, a warping loss \(\mathcal{L}_{warp}\left(I_0, I_1\right)\) is used, which ensures the quality of computed optical flow. The warping loss is defined based on the warped images $I_1 \circ \phi_{1 \rightarrow 0}^\theta$ and $I_0 \circ \phi_{0 \rightarrow 1}^\theta$, which corresponds to \(I_0\) and \(I_1\), respectively. \(L_{warp}\) can be expressed as
\begin{align}
    \mathcal{L}_{warp}^{\theta}(I_0, &I_1) = \; \mathcal{L}_{smth}(\phi_{0 \rightarrow 1}^\theta) + \mathcal{L}_{image}(I_1, I_0 \circ \phi_{0 \rightarrow 1}^\theta) \notag \\
    + \; &\mathcal{L}_{smth}(\phi_{1 \rightarrow 0}^\theta) + \mathcal{L}_{image}(I_0, I_1 \circ \phi_{1 \rightarrow 0}^\theta),
\end{align}
where \(\mathcal{L}_{smth}\) is a smoothness term that promotes similar flow values among neighboring voxels, and \(\mathcal{L}_{image}\) ensures alignment between two images. Typically, we utilize the sum of normalized cross-correlation (NCC)~\cite{voxelmorph2019} and Charbonnier~\cite{charbonnier1994two} losses as the \(\mathcal{L}_{image}\), since NCC has extensively used is 3D medical flow calculation works~\cite{voxelmorph2019, Zhao2019ICCV, shu2021ulaenet}, and Charbonnier loss is a common choice in previous VFI works~\cite{Zhao2019ICCV, shu2021ulaenet, park2023biformer, jin2023unified, kim2023event, wei2023mpvf}. The losses are defined as:
\begin{align}
    &\mathcal{L}_{smth}(\phi) = \; \lVert \nabla \phi \rVert_2 \\
    \mathcal{L}_{image}(I, \hat I) &= \; - NCC(I, \hat{I}) + \sqrt{(I - \hat I)^2 + \epsilon^2}, \label{eq:l_image}
\end{align}
where $\nabla \phi$ denotes the flow gradient, and $\epsilon$ represents a small constant. 

The fully learned flow calculation model, denoted as \(\phi^{\theta^*}_{0 \rightarrow 1}\), is earned by minimizing the warping loss \(\mathcal{L}_{warp}^{\theta}(I_0, I_1)\) with respect to the model parameter \(\theta\). The calculated flow can be formulated as:
\begin{align}\label{posthoc_flow1}
\phi^{\theta^*}_{0 \rightarrow 1} \quad \textup{s.t.} \quad \theta^* :=  \underset{\theta}{\arg\min}\sum_{(I_0, I_1) \in \mathcal{D}} \mathcal{L}_{warp}^{\theta}\left(I_0, I_1\right),
\end{align}
where \(\mathcal{D}\) indicates the training set containing the pairs of \(I_0\) and \(I_1\).

\subsection{Previous unsupervised VFI approaches}
\label{subsec:baseline approach}
\paragraph{Methodology.} If the flow from \(I_0\) to the intermediate target sample \(I_t\) can be ideally acquired as \(\phi_{0 \rightarrow t}\), the corresponding \(\hat I_t\) can also be obtained (i.e., $\hat{I_t} = I_0 \circ \phi_{0 \rightarrow t}$). To obtain this flow, current approaches~\cite{kim2022diffusion, voxelmorph2019} approximate $\phi_{0 \rightarrow t}$ as the following linear interpolation of the flow or latent vector: 
\begin{align}\label{baseline_flow1}
\phi^{\theta^*}_{0 \rightarrow t} := t \cdot \phi^{\theta^*}_{0 \rightarrow 1} \quad \textup{or} \quad \phi^{\theta^*}_{0 \rightarrow t} := \phi^{t \cdot \theta^*}_{0 \rightarrow 1},
\end{align}
where \(t \cdot \theta\) indicates the linear multiplication of latent vector. Therefore, the target $I_t$ can be obtained by approximating it as \(\hat I_t := I_0 \circ \phi^{\theta^*}_{0 \rightarrow t}\).

\paragraph{Limitations.} As detailed in the latter part of~\cref{subsec:related work video interpolation}, existing post-hoc linear interpolation approaches encounter two major challenges: firstly, they are prone to spatial distortion since the underlying network $\mathcal{F}^{\theta^*}$ that $\hat I_t$ relies on does not account for the structural smoothness between two samples during network training; and secondly, they often suffer from artifacts resulting from the warping procedure. Moreover, the methods heavily rely on post-hoc linear multiplication, leading to potential overfitting to the linear assumption. In other words, these methods assume that the structures within a given 4D medical image move only in a linear direction, and the magnitude of this movement is linearly proportional to time. 

\section{Method}
\label{sec:method}
We introduce our Unsupervised Volumetric Interpolation Network, referred to as UVI-Net. The network first generates intermediate images and then employs cycle consistency constraints to reconstruct authentic images from these synthesized ones. In \cref{subsec:our approach}, we provide an overview and a detailed presentation of our method. The training and inference procedures are outlined in \cref{subsec:training} and \cref{subsec:inference}, respectively. Additionally, in \cref{subsec:instance-specific-optimization}, we introduce an instance-specific optimization method to further enhance our model's performance.

\begin{figure}
  \centering
  \includegraphics[width=0.37\textwidth]{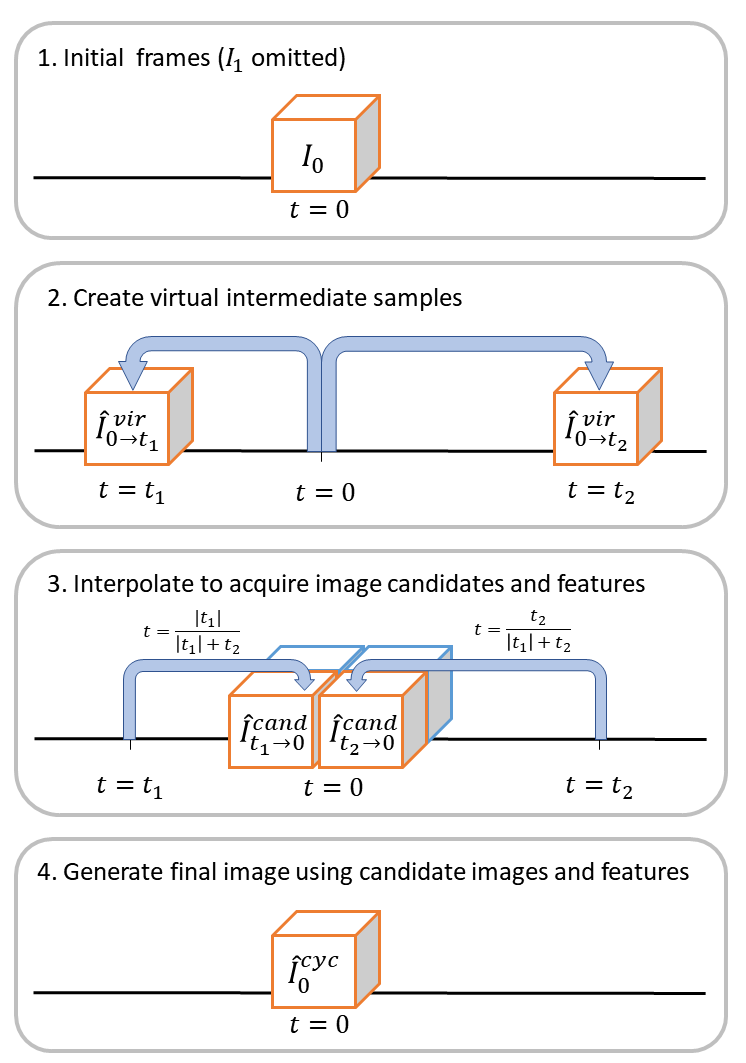}
  \vspace{-0.3cm}
  \caption{An overview of time-domain cycle consistency constraint. This image illustrates the process of generating \(\hat I_0^{cyc}\). (1) \(I_0\) and \(I_1\) are given two input frames, with \(I_1\) ommited for sake of readability. (2) We first generate virtual intermediate frames, and (3) subsequently generate back the frames with multi-resolution features (denoted as blue cubics). (4) The resulting reconstructed images \(\hat I_0^{cyc}\) must match the original input frame, \(I_0\).}
  \label{fig:cycle_consistency}
  \vspace{-0.3cm}
\end{figure}

\subsection{Methodology overview}
\label{subsec:our approach}
To achieve a result exhibiting improved smoothness for the intermediate sample derived from the network, it is imperative for the network to access the pertinent information to the intermediate sample during the learning process. In light of this, we propose the cyclic structure model, which first generates the intermediate images and reconstructs them back to the two given input images. To ensure consistency and coherence in the generated images, we impose constraints of cycle consistency \(\mathcal{L}_{cyc}\) between the reconstructed samples, denoted as \(\hat I_0^{cyc}, \hat I_1^{cyc}\), and the corresponding original samples \(I_0, I_1\). The flow of the intermediate frame is then estimated using our flow calculation model with the parameter \(\theta^*\) as follows:
\begin{align}
    \phi^{\theta^*}_{0 \rightarrow t} &:= t \cdot \phi^{\theta^*}_{0 \rightarrow 1} \quad  \textup{s.t.}\label{porposed_flow1} \\
    \theta^* := \underset{\theta}{\arg\min}\,&\underset{\omega, \psi}{\min}\sum_{(I_0, I_1) \in \mathcal{D}}\mathcal{L}_{warp}^{\theta}\left(I_0, I_1\right) \notag \\ + \mathcal{L}_{cyc}^{(\theta,\omega,\psi)}&\left(I_0, \hat I_{0}^{cyc}\right) + \mathcal{L}_{cyc}^{(\theta,\omega, \psi)}\left(I_1, \hat I_{1}^{cyc}\right), \label{eq:theta_cvin}
\end{align}
where $\theta$, $\omega$, and $\psi$ indicate the parameters for the flow calculation, feature extraction, and reconstruction models, which will be described in the below sections. 

Unlike the current approach in \cref{posthoc_flow1}, we allow the network to access intermediate samples and update them in its training, as described in~\cref{eq:theta_cvin}, resulting in improved natural voxels. We first explain the process of obtaining \(\hat I^{cyc}\) and provide a detailed explanation of \(\mathcal{L}_{cyc}\) in the following sections.

\subsection{Training}
\label{subsec:training}
The overall acquire procedure of \(\hat I_0^{cyc}\) and \(\hat I_1^{cyc}\) is illustrated in~\cref{fig:cycle_consistency}. First, we generate multiple virtual intermediate samples (see Step 2 in~\cref{fig:cycle_consistency}) by randomly sampling values of \(t_1, t_2\), and \(t_3\) as below.
\begin{align}
    \hat I_{t_1}^{vir} &:= I_0 \circ \left ( t_1 \cdot \phi^\theta_{0 \rightarrow 1} \right ) \qquad \quad -0.5 \le t_1 \le 0 
    \makebox[0cm]{} \label{eq:it11st} \\
    \hat I_{t_2}^{vir} &:= 
    \begin{cases}
        I_0 \circ \left ( t_2 \cdot \phi^\theta_{0 \rightarrow 1} \right ) \qquad \quad \;\; 0 \le t_2 \le 0.5 \\ 
        I_1 \circ \left ( (1-t_2) \cdot \phi^\theta_{1 \rightarrow 0} \right ) \quad 0.5 \le t_2 \le 1 
    \end{cases} \\
    \hat I_{t_3}^{vir} &:= I_1 \circ \left ( (1-t_3) \cdot \phi^\theta_{1 \rightarrow 0} \right ) \qquad 1 \le t_3 \le 1.5 \label{eq:it31st}
\end{align}

Since \(\hat I_{t_1}^{vir}\) and \(\hat I_{t_3}^{vir}\) are generated outside the time range between the two frames, we limit the maximum time offset to 0.5 to mitigate the occurrence of artifacts. When generating the \( \hat I_{t_2}^{vir}\), a synthesized image between the two input images, we adopt the result created from the image—either $I_0$ or $I_1$—that is closer to the reference point $t_2$, to preserve the properties of the real image maximally.

Next, we interpolate the generated intermediate samples (see Step 3 in \cref{fig:cycle_consistency}) to acquire the \(I_0\) and \(I_1\)'s candidates as follows:
\begin{align}
    \hat I_{t_1 \rightarrow 0}^{cand} := \hat I^{vir}_{t_1} \circ \left ( \frac{-t_1}{t_2 - t_1} \cdot \phi^{\theta}_{t_1 \rightarrow t_2} \right ), \label{eq:hati0i1_start} \\
    \hat I_{t_2 \rightarrow 0}^{cand} := \hat I^{vir}_{t_2} \circ \left ( \frac{t_2}{t_2 - t_1} \cdot \phi^{\theta}_{t_2 \rightarrow t_1} \right ), \\
    \hat I_{t_2 \rightarrow 1}^{cand} := \hat I^{vir}_{t_1} \circ \left ( \frac{1-t_2}{t_3 - t_2} \cdot \phi^{\theta}_{t_2 \rightarrow t_3} \right ), \\
    \hat I_{t_3 \rightarrow 1}^{cand} := \hat I^{vir}_{t_3} \circ \left ( \frac{t_3-1}{t_3 - t_2} \cdot \phi^{\theta}_{t_3 \rightarrow t_2} \right ). \label{eq:hati0i1_end}
\end{align} 

\begin{figure}[t]
  \centering
  \includegraphics[width=0.3\textwidth]{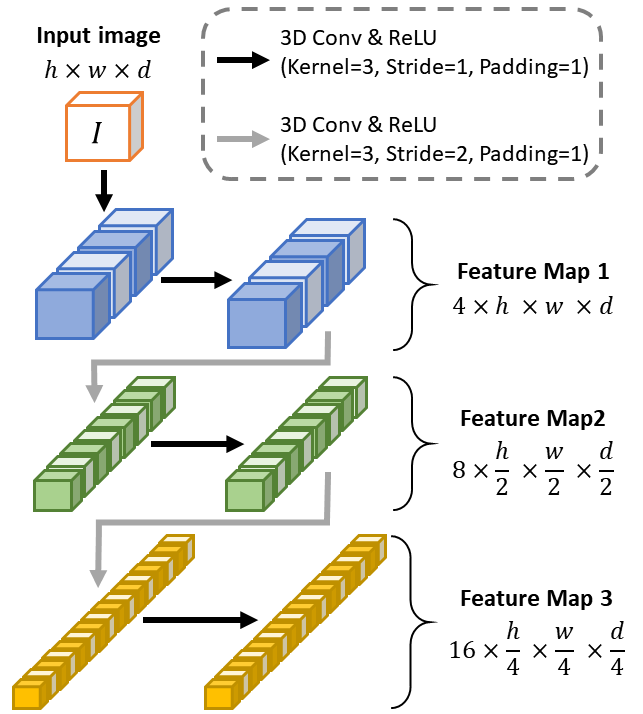}
  \vspace{-0.2cm}
  \caption{Architecture of the feature extractor module based on 3D Convolutional Neural Network (CNN). $h$, $w$, and $d$ are the input image's height, width, and depth, respectively.}
  \label{fig:feature_extract}
  \vspace{-0.3cm}
\end{figure}

While warping the virtual frames, we simultaneously warp the feature space of the frames across multiple resolutions, obtaining a set of warped feature maps: 
\(\mathcal{S}_{t_1 \rightarrow 0}\), \(\mathcal{S}_{t_2\rightarrow 0}\), \(\mathcal{S}_{t_2 \rightarrow 1}\), and \(\mathcal{S}_{t_3 \rightarrow 1}\). Specifically, following the architecture of our feature extractor as shown in~\cref{fig:feature_extract}, we extract feature maps resized to 1, 0.5, and 0.25 times their original size. Then, using the same optical flow as described in \cref{eq:hati0i1_start} to (\ref{eq:hati0i1_end}) (or downscaled as necessary), we obtain the final warped feature maps. This method enhances the reconstruction model's ability to make more accurate predictions by providing access to both voxel and feature information. Furthermore, we extract image representations at various levels, which have proven effective in previous research on video-related tasks \cite{niklaus2020softmax, jin2023unified, karim2023med}.

\begin{figure*}[t]
  \centering
  \includegraphics[width=0.8\textwidth]{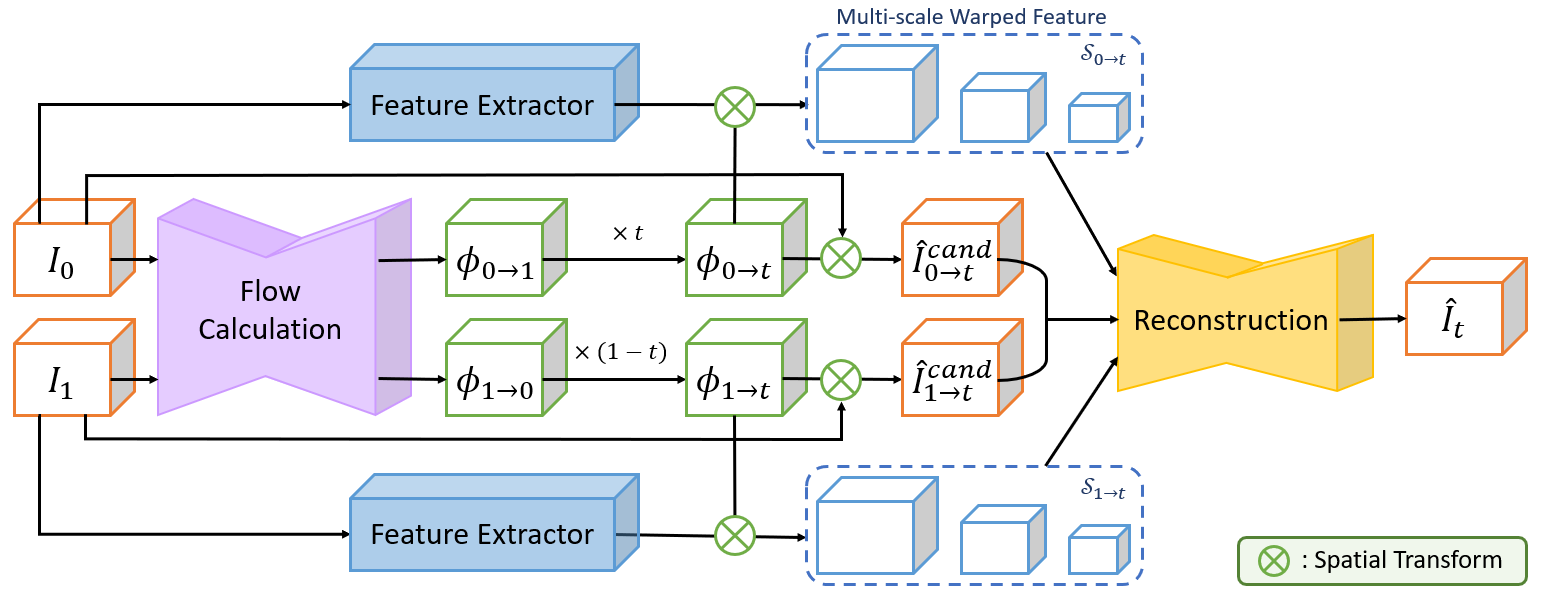}
  \vspace{-0.15cm}
  \caption{Schematic overview of our entire inference process. Starting with two input frames, \(I_0\) and \(I_1\), we input the frames into the flow calculation model to obtain the approximated flow fields \(\phi_{0 \rightarrow t}\) and \(\phi_{1 \rightarrow t}\). We then warp the two frames using the obtained flow field, and similarly warp the multi-scale voxelwise features. Finally, we refine the distance-inversely weighted added image considering the information from multi-scale features, resulting in the final interpolated frame \(\hat I_t\).}
  \label{fig:main_figure}
  \vspace{-0.4cm}
\end{figure*}

Using these warped images and features, we obtain the predictions $\hat I_{0}^{cyc}$ and $\hat I_{1}^{cyc}$ using the reconstruction model \(\mathcal{R}^\psi\) (see Step 4 in~\cref{fig:cycle_consistency}). The model takes the distance-based weighted sum images and warped feature map sets, and reconstructs the original frames through residual corrections. Each element of the input feature map sets is fed into individual encoder layers of the reconstruction model and concatenated channel-wise. The procedure of the reconstruction model can be written as:
\begin{align}
    \hat I_{0}^{cyc} := \; &\mathcal{R}^\psi(\, \hat I_{t_1 \rightarrow 0}^{cand} \oplus \hat I_{t_2 \rightarrow 0}^{cand},\, \mathcal{S}_{t_1 \rightarrow 0},\, \mathcal{S}_{t_2\rightarrow 0} \, ),  \\
    \hat I_{1}^{cyc} := \; &\mathcal{R}^\psi(\, \hat I_{t_2 \rightarrow 1}^{cand} \oplus \hat I_{t_3 \rightarrow 1}^{cand},\, \mathcal{S}_{t_2 \rightarrow 1},\, \mathcal{S}_{t_3\rightarrow 1} \, ),
\end{align}
where \(\oplus\) indicates distance-based addition.

With the reconstructed images \(\hat I_0^{cyc}\) and \(\hat I_1^{cyc}\), we can introduce the cycle consistency loss. Our cycle-consistent framework reconstructs real images from the generated intermediate images, thereby enhancing the smoothness of the interpolated images. Without time notation for clarity, consider the reconstructed image (\(\hat I^{cyc}\)) and corresponding real image (\(I\)). The cycle consistency loss is defined as:
\begin{align}
    \mathcal{L}_{cyc}^{(\theta,\omega,\psi)}(I, \hat I^{cyc}) &= \mathcal{L}_{image}(I, \hat I^{cyc}) + \mathcal{L}_{reg}(\mathcal{R}^\psi), 
\end{align}
where
\(\mathcal{L}_{image}\) follows~\cref{eq:l_image}, and \(\mathcal{L}_{reg}\) acts as an L1 regularization term applied to the predicted residual of the reconstruction model. This term helps control excessive modification during the reconstruction process. 

In essence, even without any intermediate frames, we utilize the given authentic frames as pseudo supervision for the intermediate frame, facilitated by the initially generated virtual intermediate samples \(\hat I^{vir}\). By incorporating a cycle consistency constraint between the reconstructed and original authentic images, our approach enhances spatial continuity between the two images and generates high-quality virtual intermediate samples.
    
\input{table/main_table_final}

\subsection{Inference}
\label{subsec:inference}
We illustrate the overall inference procedure of UVI-Net in~\cref{fig:main_figure}. First, we obtain two optical flow \(\phi_{0 \rightarrow 1}^{\theta^*}\) and \(\phi_{1 \rightarrow 0}^{\theta^*}\), where \(\theta^*\) follows~\cref{eq:theta_cvin}. Next, we attain two \(I_t\)'s candidate as follows:
\begin{align}
    \hat I_{0 \rightarrow t}^{cand} := \; &I_0 \circ \phi_{0 \rightarrow t} = I_0 \circ \left( t \cdot \phi_{0 \rightarrow 1}^{\theta^*} \right)\\
    \hat I_{1 \rightarrow t}^{cand} := \; &I_1 \circ \phi_{1 \rightarrow t} = I_1 \circ \left( (1-t) \cdot \phi_{1 \rightarrow 0}^{\theta^*} \right).
\end{align}
Finally, by reconstructing the final image with the two candidates considering the temporal distance, we derive \(\hat I_{t}\) as
\begin{align}
    \hat I_t := \; & \mathcal{R}^\psi(\, \hat I_{0 \rightarrow t}^{cand} \oplus \hat I_{1 \rightarrow t}^{cand},\, \mathcal{S}_{0 \rightarrow t},\, \mathcal{S}_{1 \rightarrow t} \, ) ,
\end{align}
where $\mathcal{S}_{0 \rightarrow t}$ and $\mathcal{S}_{1 \rightarrow t}$ are warped feature map sets from $I_0$ and $I_1$, respectively. 
Remarkably, while baseline approaches can only use one of \(\hat I_{0 \rightarrow t}\) and \(\hat I_{1 \rightarrow t}\), we can engage both information and make to be symmetric even the order of \(I_0\) and \(I_1\) is switched.

\subsection{Instance-Specific Optimization}
\label{subsec:instance-specific-optimization}
Instance-specific optimization is a technique used to enhance the final performance by fine-tuning models for each test sample. This approach was introduced by \citet{balakrishnan2018unsupervised} within the unsupervised medical image warping domain. Despite our work being in a different task, this strategy remains applicable. Utilizing a model weight pre-trained on the training data, we fine-tune the model for a relatively small number of epochs on each test data. Such an adaptive approach is particularly beneficial in medical imaging, allowing for more personalized and accurate frame interpolation tailored to individual scans. 

\section{Experiments}
\label{sec:experiment}
This section describes the benchmark datasets for 4D medical imaging used in this study in~\cref{subsec:datasets}. Next,~\cref{subsec:experimental settings} outlines some settings, including training details and metrics for performance evaluation. The results are comprehensively presented in ~\cref{subsec:results}, highlighting our method's effectiveness and efficiency.

\subsection{Datasets}
\label{subsec:datasets}
To evaluate the performance of image interpolation, two 4D image datasets are used, each for the heart and lung. The ACDC cardiac dataset~\cite{bernard2018deep} consists of 100 4D temporal cardiac MRI images. End-diastolic and end-systolic phase images are used as the start and end images, respectively. The initial 90 alphabetically sorted samples form the training set, with the remaining used for the test set. The 4D-Lung dataset~\cite{hugo2017longitudinal} consists of 82 chest CT scans for radiotherapy planning from 20 lung cancer patients. In each 4D-CT study, the end-inspiratory (0\% phase) and end-expiratory (50\% phase) phase scans are set as the initial and final images, respectively.  The first 68 CT scans from 18 patients in the dataset are included in the training set. For additional information for dataset, please refer to~\cref{sec:supple 4d dataset}.

\begin{figure*}[t]
     \centering
     \begin{subfigure}[b]{\textwidth}
         \centering
         \includegraphics[width=.98\textwidth]{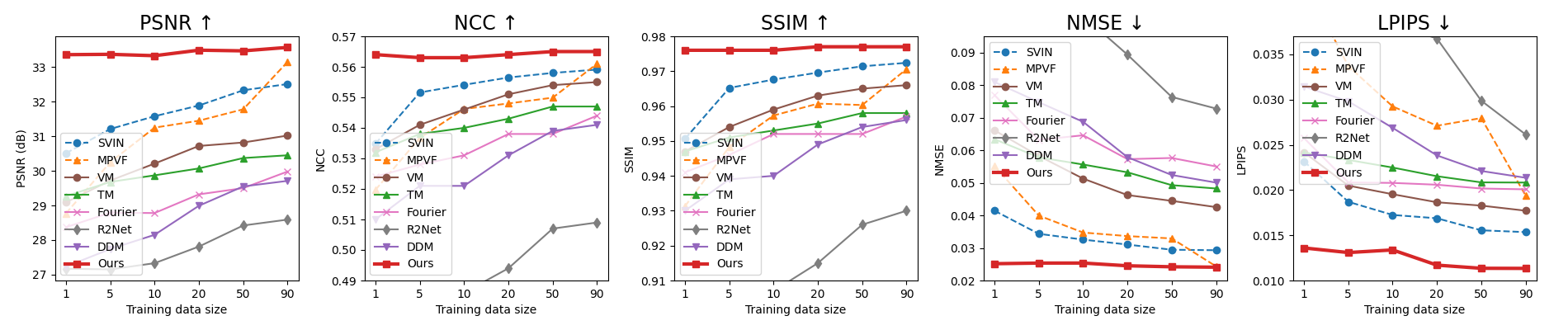}
         \caption{Performance on the cardiac dataset according to training size.}
         \label{fig:ablation cardiac}
     \end{subfigure} \\
     \hfill
     \begin{subfigure}[b]{\textwidth}
         \centering
         \includegraphics[width=.98\textwidth]{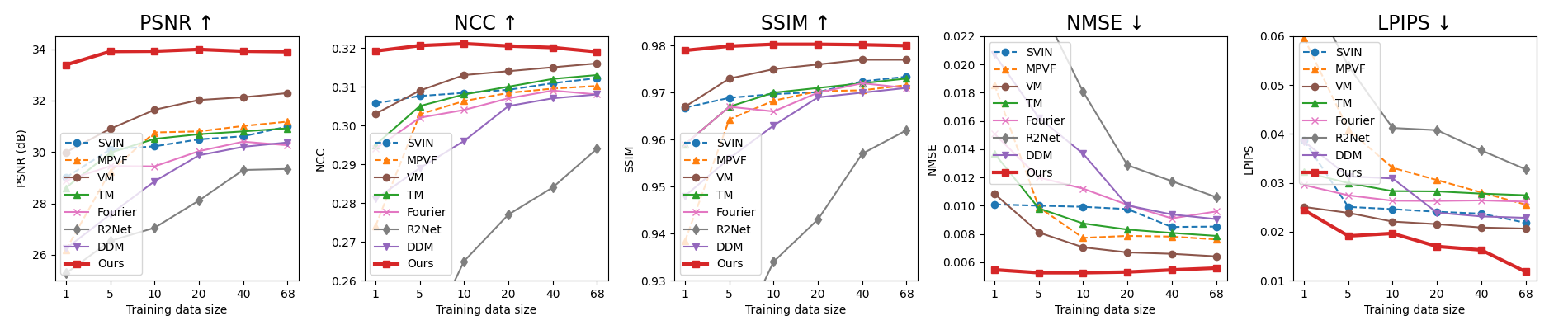}
         \caption{Performance on the lung dataset according to training size.}
         \label{fig:ablation lung}
     \end{subfigure}
    \caption{Performance trends based on the size of the training datasets. The dashed line represents a supervised setting. As depicted in this figure, we observe that the performance gap between our model and the baselines increases regardless of whether the setting is supervised or not, and irrespective of the dataset type. This demonstrates our model's robustness, particularly in addressing data scarcity issues common in the medical domain.}
    \label{fig:ablation}
    \vspace{-0.1cm}
\end{figure*}

\subsection{Experimental settings}
\label{subsec:experimental settings}
\subsubsection{Baselines}
For comparison with our proposed methods, six models are included as the baselines. VoxelMorph (VM)~\cite{balakrishnan2018unsupervised}, TransMorph (TM)~\cite{chen2022transmorph}, Fourier-Net+~\cite{jia2023fourier} and R2Net~\cite{joshi2023r2net} are first initially trained with the provided dataset to calculate optical flow. Interpolated images are then obtained by linear scaling the optical flow, i.e., \(t \cdot \phi^{\theta^*}_{0 \rightarrow t}\). Diffusion Deformable Model (DDM)~\cite{kim2022diffusion} also uses dataset training but interpolates by scaling the latent vector, i.e., \(\phi^{t \cdot \theta^*}_{0 \rightarrow t}\). For IDIR~\cite{wolterink2022implicit}, it is crucial to clarify that it requires individual training for each target registration pair, leading to limited generalization, whereas our method is trained using a distinct training set and subsequently applied for inference on the target pairs. We also compared the results of our model with two supervised methods proposed for video interpolation on 4D medical images: SVIN \cite{guo2020spatiotemporal} and MPVF \cite{wei2023mpvf}.  Detailed information about the baseline models is in~\cref{sec:supple baseline}.

\subsubsection{Evaluation metrics}
To evaluate the similarity between the predicted and ground truth images, metrics including PSNR (Peak Signal-to-Noise Ratio)~\cite{dosselmann2005psnr}, NCC (Normalized Cross Correlation), SSIM (Structural Similarity Index Measure)~\cite{ssim2004}, NMSE (Normalized Mean Squared Error) and LPIPS (Learned Perceptual Image Patch Similarity)~\cite{zhang2018unreasonable} are used. Since LPIPS is available only for 2D, it was averaged across slices along the x, y, and z axes. Each metric represents the voxel-wise similarity, correlation, structural similarity, reconstruction error, and perceptual similarity between the synthesized and authentic images.

\vspace{-0.3cm}
\subsubsection{Training details}
\vspace{-0.1cm}
For the flow calculation model, we employed the network designed in VoxelMorph~\cite{balakrishnan2018unsupervised}. As for the reconstruction model, we used a small size of 3D-UNet. The detailed configuration of the network and more details are described in~\cref{sec:supple cvin details}. The proposed method was implemented with PyTorch~\cite{NEURIPS2019_9015} using an NVIDIA Tesla V100 GPU. The training process takes approximately 4 hours for the cardiac dataset and 8 hours for the lung dataset, respectively. Instance-specific optimization took about 1.12 minutes per sample for ACDC and 3.13 minutes for 4D-Lung.

\vspace{+0.2cm}
\subsection{Results}
\label{subsec:results}
\subsubsection{Interpolation Result}
The performance of interpolation compared to unsupervised and supervised methods is shown in~\cref{tab:interpolation result}. Our method consistently demonstrates superior performance among all the models, outperforming others with a significant margin in every evaluation metric. This trend is observed across both heart and lung datasets, even in the absence of instance-specific optimization.

It is important to note that our approach surpasses IDIR~\cite{wolterink2022implicit}, serving as a rigorous comparison baseline for our method due to IDIR's test set-specific optimization. The core methodology behind IDIR undergoes unique adaptation for each test set pair, which involves retraining for every new instance. While this strategy enables IDIR to tailor its performance to each dataset, it restricts its practical applicability. Nevertheless, our method demonstrates substantial superiority over IDIR in terms of performance.

\begin{figure*}[t]
  \centering
  \includegraphics[width=0.87\textwidth]{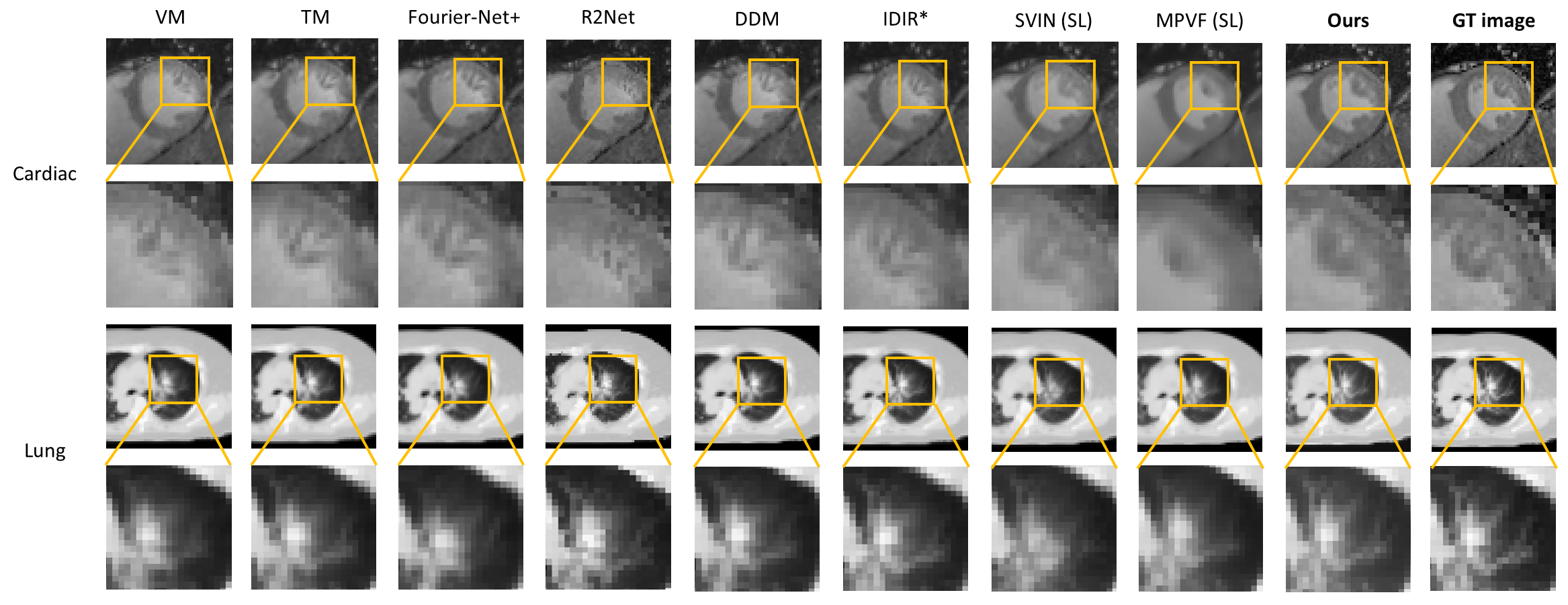}
  \vspace{-0.2cm}
  \caption{Visualization examples from 4D cardiac and lung datasets. The model marked with an `*' is trained exclusively on the test set, while models marked with `(SL)' are trained using supervised learning. Our method generates intermediate frames that are not only visually appealing but also precise, successfully retaining fine details and maintaining the structural integrity of the original images.}
  \label{fig:comparison}
  \vspace{-0.3cm}
\end{figure*}

\vspace{-0.2cm}
\paragraph{Supervised Models.} Notably, our approach also surpasses supervised methods. An interesting observation is the varying performance of these supervised models across different datasets. As detailed in~\cref{tab:dataset comparison}, the ACDC dataset contains significantly more frames compared to the 4D-Lung dataset. This discrepancy implies that the 4D-Lung dataset experiences limitations in terms of supervision quality. Therefore, the performance gap is more pronounced in the lung dataset, underscoring a critical insight: supervised models tend to underperform with limited supervision from intermediate frames. This pattern reaffirms the importance of our method's ability to achieve high accuracy in scenarios with constrained supervision, highlighting its robustness and effectiveness in 4D medical VFI tasks.

\subsubsection{Effect of training dataset size}
\cref{fig:ablation} illustrates the interpolation performance based on the number of training samples. With the test sets remaining fixed, the sizes of the training sets are reduced from their full down to one. Compared to the other five unsupervised baselines (VM, TM, Fourier-Net+, R2Net, and DDM), our method consistently exhibits superior performance across varying training set sizes, with performance gaps widening as the dataset size decreases. Remarkably, even with a minimal size comprising \textit{only one sample}, our approach frequently outperforms the baselines that utilize the maximum training set size. It should be noted that IDIR is not included in this comparison, as it does not follow a traditional training process on a training set. For the two supervised baseline models (SVIN, MPVF), the performance also diminishes as the number of samples for supervision decreases, leading to an increasing performance gap between them and our model. Our consistent performance in scenarios with small datasets underscores the strengths of our approach in mitigating the challenges posed by data scarcity in the medical field.

\subsubsection{Qualtitative analysis}
The comparison of qualitative results between interpolation methods is shown in~\cref{fig:comparison}. Our method consistently produces visually appealing and accurate intermediate frames, capturing fine details and preserving the structural integrity of the original images.

\subsubsection{Downstream task}
We also demonstrate that our interpolation method can be applied to downstream tasks. Specifically, we tested its effectiveness on segmentation data, which is relatively complex, demonstrating our approach's potential for augmenting 3D medical datasets. Details of the experimental setup and performance can be found in~\cref{sec: augmentation}.

\subsubsection{Additional experiments} 
Additional experiments, including ablation studies and further qualitative results, are detailed in~\cref{sec: supple additional experiment}. Moreover, we have analyzed the results of extrapolation to ensure that generated images during the training process do not exhibit any unnatural changes or issues.

\section{Conclusion}
\label{sec:conclusion}
Our framework, UVI-Net, effectively tackles the challenge of generating intermediate frames for 4D medical images through unsupervised volumetric interpolation. By leveraging pseudo supervision within a cyclic structure, our method ensures spatial continuity between the generated intermediate and real images. Experimental results on benchmark datasets validate the efficacy of our approach, revealing substantial improvements in intermediate frame quality across various evaluation metrics, surpassing both unsupervised and supervised baselines. Furthermore, our method has demonstrated robustness not only in situations of frame scarcity but also in data scarcity contexts. Ultimately, this study underscores the promise of unsupervised 3D flow-based interpolation and opens new avenues for research and development in the field of medical imaging.

\paragraph{Acknowledgement}
This study was supported by Institute for Information \&
communications Technology Promotion (IITP) grant funded by
the Korea government (MSIT) (No.2019-0-00075 Artificial Intelligence Graduate School Program (KAIST)) and Medical Scientist Training Program from the Ministry of Science \& ICT of Korea.

\clearpage
{
    \small
    \bibliographystyle{ieeenat_fullname}
    \bibliography{main}
}

\input{supplementary}

\end{document}

%% file: preamble.tex
\usepackage[dvipsnames]{xcolor}


%% file: table/compare_dataset.tex
        

\begin{table}[t]
    \centering
    \begin{adjustbox}{width=.37\textwidth}
    \begin{tabular}{clc}
        \toprule 
        \textbf{Data Type} & \textbf{Name} & \textbf{\# of Total Inter} \\
        \midrule
        \multirow{5}{*}{2D Natural} & 
        UCF101~\cite{soomro2012ucf101} & 2,374,290 \\
        & X4K1000FPS~\cite{sim2021xvfi} & 277,704 \\
        & Adobe240-fps~\cite{su2017deep} & 79,768\\
        & Vimeo90K~\cite{xue2019video} & 73,171 \\
        & ATD-12K~\cite{siyao2021deep}  & 12,000\\
        \cdashline{1-3}
        \multirow{2}{*}{3D Medical} & ACDC~\cite{bernard2018deep} & 2,556 \\
         & 4D-Lung~\cite{hugo2017longitudinal} & 648 \\
        \bottomrule
    \end{tabular}
    \end{adjustbox}
    \vspace{-0.1cm}
    \caption{Comparison of representative 2D VFI datasets with 3D medical VFI datasets in our study. The last column indicates the total number of intermediate frames, representing the sum of intermediate frame counts across each dataset.}
    \label{tab:dataset comparison}
    \vspace{-0.5cm}
\end{table}

%% file: table/main_table_final.tex
\begin{table*}[t]
  \centering
  \begin{adjustbox}{width=.85\linewidth}
  \begin{tabular}{lclccccc}
    \toprule
    \textbf{Dataset} &  \textbf{Supervised}&\textbf{Method} & PSNR \(\uparrow\) & NCC \(\uparrow\) & SSIM \(\uparrow\) & NMSE \(\downarrow\) & LPIPS \(\downarrow\)\\
    \midrule
    \multirow{10}{*}{\textbf{Cardiac}}& \multirow{2}{*}{\ding{51}}& SVIN~\cite{guo2020spatiotemporal} & 32.51 \scriptsize{\(\pm 0.254\)}& 0.559 \scriptsize{\(\pm 0.007\)}& 0.972 \scriptsize{\(\pm 0.001\)}& 2.930 \scriptsize{\(\pm 0.155\)}&1.535 \scriptsize{\(\pm 0.043\)}\\
 & & MPVF~\cite{wei2023mpvf} & 33.15 \scriptsize{\(\pm 0.238\)}& 0.561 \scriptsize{\(\pm 0.006\)}& 0.971 \scriptsize{\(\pm 0.001\)}& 2.435 \scriptsize{\(\pm 0.133\)}&1.941 \scriptsize{\(\pm 0.055\)}\\
    \cdashline{2-8}&  \multirow{8}{*}{\ding{55}}&VM~\cite{voxelmorph2019}  & 31.02 \scriptsize{\(\pm 0.272\)}& 0.555 \scriptsize{\(\pm 0.006\)}& 0.966 \scriptsize{\(\pm 0.002\)}& 4.254 \scriptsize{\(\pm 0.261\)}& 1.772 \scriptsize{\(\pm 0.064\)}\\
    &  &TM~\cite{chen2022transmorph} & 30.45 \scriptsize{\(\pm 0.280\)}& 0.547 \scriptsize{\(\pm 0.006\)}& 0.958 \scriptsize{\(\pm 0.002\)}& 4.826 \scriptsize{\(\pm 0.278\)}& 2.083 \scriptsize{\(\pm 0.078\)}\\
     & & Fourier-Net+~\cite{jia2023fourier}& 29.98 \scriptsize{\(\pm 0.287\)}& 0.544 \scriptsize{\(\pm 0.006\)}& 0.957 \scriptsize{\(\pm 0.002\)}& 5.503 \scriptsize{\(\pm 0.314\)}&2.008  \scriptsize{\(\pm 0.077\)}\\
     & & R2Net~\cite{joshi2023r2net}& 28.59 \scriptsize{\(\pm 0.278\)}& 0.509 \scriptsize{\(\pm 0.007\)}& 0.930 \scriptsize{\(\pm 0.003\)}& 7.281 \scriptsize{\(\pm 0.329\)}&3.482 \scriptsize{\(\pm 0.138\)}\\
    &  &DDM~\cite{kim2022diffusion} & 29.71 \scriptsize{\(\pm 0.221\)}& 0.541 \scriptsize{\(\pm 0.006\)}& 0.956 \scriptsize{\(\pm 0.002\)}& 5.007 \scriptsize{\(\pm 0.239\)}& 2.136 \scriptsize{\(\pm 0.066\)}\\
     & & IDIR*~\cite{wolterink2022implicit}& 31.56 \scriptsize{\(\pm 0.275\)}& 0.557 \scriptsize{\(\pm 0.006\)}& 0.968 \scriptsize{\(\pm 0.001\)}& 3.806 \scriptsize{\(\pm 0.249\)}&1.675 \scriptsize{\(\pm 0.061\)}\\
    &  & \textbf{Ours (w/o inst opt.)} & \underline{33.57} \scriptsize{\(\pm 0.275\)}& \textbf{0.565} \scriptsize{\(\pm 0.007\)}& \underline{0.977} \scriptsize{\(\pm 0.001\)}& \underline{2.409} \scriptsize{\(\pm 0.159\)}& \underline{1.134} \scriptsize{\(\pm 0.044\)}\\
    &  & \textbf{Ours (w/ inst opt.)} & \textbf{33.59} \scriptsize{\(\pm 0.268\)} & \textbf{0.565} \scriptsize{\(\pm 0.007\)} & \textbf{0.978} \scriptsize{\(\pm 0.001\)} & \textbf{2.384} \scriptsize{\(\pm 0.157\)} & \textbf{1.066} \scriptsize{\(\pm 0.041\)} \\
 
    \midrule
     \multirow{10}{*}{\textbf{Lung}} & \multirow{2}{*}{\ding{51}}& SVIN~\cite{guo2020spatiotemporal} & 30.99 \scriptsize{\(\pm 0.309\)}& 0.312 \scriptsize{\(\pm 0.002\)}& 0.973 \scriptsize{\(\pm 0.002\)}& 0.852 \scriptsize{\(\pm 0.063\)}& 2.182 \scriptsize{\(\pm 0.093\)}\\
     & & MPVF~\cite{wei2023mpvf} & 31.18 \scriptsize{\(\pm 0.344\)}& 0.310 \scriptsize{\(\pm 0.003\)}& 0.972 \scriptsize{\(\pm 0.002\)}& 0.761 \scriptsize{\(\pm 0.075\)}& 2.554 \scriptsize{\(\pm 0.092\)}\\
    \cdashline{2-8}&  \multirow{8}{*}{\ding{55}}&VM~\cite{voxelmorph2019}  & 32.29 \scriptsize{\(\pm 0.314\)}& 0.316 \scriptsize{\(\pm 0.002\)}& 0.977 \scriptsize{\(\pm 0.001\)}& 0.641 \scriptsize{\(\pm 0.052\)}& 2.063 \scriptsize{\(\pm 0.108\)}\\
    &  &TM~\cite{chen2022transmorph} & 30.92 \scriptsize{\(\pm 0.290\)}& 0.313 \scriptsize{\(\pm 0.002\)}& 0.973 \scriptsize{\(\pm 0.001\)}& 0.786 \scriptsize{\(\pm 0.050\)}& 2.746 \scriptsize{\(\pm 0.113\)}\\
     & & Fourier-Net+~\cite{jia2023fourier} & 30.26 \scriptsize{\(\pm 0.314\)}& 0.308 \scriptsize{\(\pm 0.003\)}& 0.971 \scriptsize{\(\pm 0.002\)}& 0.959 \scriptsize{\(\pm 0.061\)}&2.615 \scriptsize{\(\pm 0.125\)}\\
     & & R2Net~\cite{joshi2023r2net} & 29.34 \scriptsize{\(\pm 0.270\)}& 0.294 \scriptsize{\(\pm 0.003\)}& 0.962 \scriptsize{\(\pm 0.002\)}& 1.061 \scriptsize{\(\pm 0.051\)}&3.277 \scriptsize{\(\pm 0.122\)}\\
    &  &DDM~\cite{kim2022diffusion} & 30.37 \scriptsize{\(\pm 0.271\)}& 0.308 \scriptsize{\(\pm 0.003\)}& 0.971 \scriptsize{\(\pm 0.002\)}& 0.905 \scriptsize{\(\pm 0.065\)}& 2.283 \scriptsize{\(\pm 0.106\)}\\
     & & IDIR*~\cite{wolterink2022implicit}& 32.91 \scriptsize{\(\pm 0.309\)}& \textbf{0.321} \scriptsize{\(\pm 0.003\)}& \textbf{0.980} \scriptsize{\(\pm 0.002\)}& 0.586 \scriptsize{\(\pm 0.055\)}&2.035 \scriptsize{\(\pm 0.112\)}\\
    & & \textbf{Ours (w/o inst opt.)} & \underline{33.90} \scriptsize{\(\pm 0.382\)}& 0.319 \scriptsize{\(\pm 0.003\)}&  \textbf{0.980} \scriptsize{\(\pm 0.002\)}& \underline{0.558} \scriptsize{\(\pm 0.055\)}& \underline{1.512} \scriptsize{\(\pm 0.112\)}\\
    & & \textbf{Ours (w/ inst opt.)} & \textbf{34.00} \scriptsize{\(\pm 0.387\)} & \underline{0.320} \scriptsize{\(\pm 0.003\)} &  \textbf{0.980} \scriptsize{\(\pm 0.002\)} & \textbf{0.552} \scriptsize{\(\pm 0.055\)} & \textbf{1.489} \scriptsize{\(\pm 0.093\)} \\

    \bottomrule
  \end{tabular}
  \end{adjustbox}
  \vspace{-0.2cm}
  \caption{Quantitative comparison of interpolation results. These metrics were evaluated after repeating each experiment three times and collecting all frames. The model marked with an `*' is trained exclusively on the test set, as it is designed for training on a single data pair only. For our model, the results with or without instance-specific optimization are both reported. The table presents both the average and standard deviation for each metric. NMSE and LPIPS values are presented in units of \(10^{-2}\). The best and second-best results for each metric are indicated with \textbf{bold} and \underline{underlined}, respectively.}
  \label{tab:interpolation result}
  \vspace{-0.4cm}
\end{table*}

%% file: supplementary.tex
\clearpage

\appendix
\appendixpage

\section{Details of 4D datasets}
\label{sec:supple 4d dataset}

\paragraph{ACDC.}
The ACDC dataset features an average of 10.02$\pm$2.20 frames between the end-systolic and end-diastolic phases in the training set, with the test set presenting an average of 8.80$\pm$2.48 frames. All cardiac MRI scans have been uniformly resized. Following this resizing process, min-max scaling is applied to ensure consistent scaling across all scans.

\paragraph{4D-Lung.}
In the case of the 4D-lung dataset, the models are trained to predict the four intermediate frames (10\%, 20\%, 30\%, 40\%) between the end-inspiratory (0\%) and end-expiratory (50\%) phases. Only CT images captured using kilovoltage energy are included in the study due to their superior image quality. Each lung CT scan is adjusted to the lung window range (-1400 to +200 Hounsfield unit)~\cite{grob2019lungwindow} and subjected to centering and min-max scaling. Subsequently, bed removal is performed using the following method: pixels exceeding a certain threshold (-500 HU in this study) are assigned a value of 1, while all other pixels are set to 0, creating a binarized map. The binarized map undergoes erosion/dilation~\cite{soille2004dilation} to identify the most prominent body contour mask. By getting the resulting body contour mask to the corresponding voxel region of the given images, a bed-removed CT image is obtained. All the lung CT images are resized to $128 \times 128 \times 128$.

\newpage
\section{Details of baseline models}
\label{sec:supple baseline}
The following three unsupervised models and two supervised models are used as the baseline models for our main result: VoxelMorph~\cite{voxelmorph2019}, TransMorph~\cite{chen2022transmorph}, Fourier-Net+~\cite{jia2023fourier}, R2Net~\cite{joshi2023r2net}, IDIR~\cite{wolterink2022implicit}, DDM~\cite{kim2022diffusion} for unsupervised models, and SVIN~\cite{guo2020spatiotemporal}, MPVF~\cite{wei2023mpvf} for supervised models. To the best of our knowledge, this selection covers the most pertinent and all current baseline models in the field, providing a comprehensive benchmark for our study.

\paragraph{Unsupervised models.} 
The VoxelMorph employs the exact same model architecture as our flow calculation model, as discussed in \cref{subsec:flow detail}. For TransMorph, we follow the TransMorph-Large framework from the original paper. In the case of Fourier-Net+, R2Net, IDIR, and DDM, we utilize the default architecture outlined in the original paper. 

\paragraph{Supervised models.}
In our study involving SVIN, we adhered to the official architecture as described in the foundational paper. For MPVF, we applied the architecture specified for the ACDC dataset, as outlined in the original publication. However, our experience with the 4D-lung dataset presented unique challenges. Despite the original study using a distinct lung preprocessing method, which resulted in larger data sizes, and reporting successful execution on a V100 GPU with 32GB of memory, our attempts to run their code on an A6000 GPU with 48GB of memory encountered memory issues. Upon contacting the authors, we learned that no official code was available for the 4D-lung dataset. Consequently, we were compelled to arbitrarily modify the model size to accommodate our 48GB memory constraint. This entailed reducing the encoder inplanes from [32, 64, 128] to [8, 16, 32], decreasing the number of ViT heads from 4 to 2, lowering the ViT num classes from 1000 to 300, and diminishing the hidden dimension from 256 to 64. Please note that although we reduced the model to fit a 48GB memory constraint, our measurements were conducted on a model size larger than the original model's 32GB specification.

\newpage
\begin{figure*}[t]
     \centering
     \begin{subfigure}[b]{\textwidth}
         \centering
         \includegraphics[width=.95\textwidth]{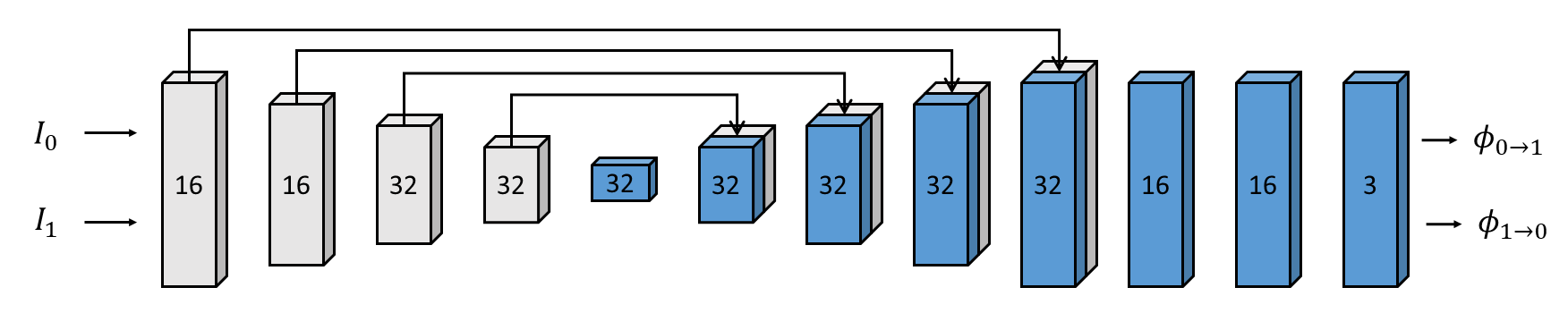}
         \caption{Architecture of flow calculation model}
         \label{fig:architecture flow}
     \end{subfigure} \\
     \hfill
     \begin{subfigure}[b]{\textwidth}
         \centering
         \includegraphics[width=.95\textwidth]{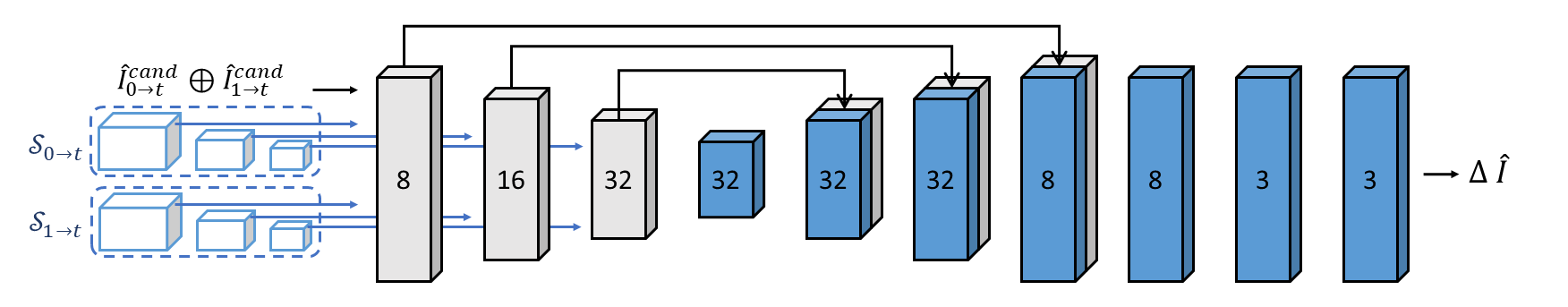}
         \caption{Architecture of reconstruction model}
         \label{fig:architecture recon}
     \end{subfigure}
    \caption{The architecture of the flow calculation and reconstruction models consisting of encoder and decoder layers. The encoder layers are represented by gray boxes, while the decoder layers are represented by blue boxes. The numbers associated with each box indicate the number of features in the corresponding convolutional filter.}
    \label{fig:architecture}
\end{figure*}

\section{UVI-Net details}
\label{sec:supple cvin details}

\subsection{Flow calculation model}
\label{subsec:flow detail}

The flow calculation model follows the network architecture illustrated in \cref{fig:architecture flow}, which is based on VoxelMorph~\cite{voxelmorph2019}. The model processes a single input by combining the images \(I_0\) and \(I_1\) into a 2-channel 3D image. Then, it outputs 3-channel 3D flows, where each channel represents the displacement along each dimension. The flow model incorporates 3D convolutions in both the encoder and decoder stages with a kernel size of 3. LeakyReLU layer with a negative slope of 0.2 follows each convolutional operation. 

In the encoder, strided convolutions with stride size 2 are utilized to reduce the spatial dimensions by half at each layer. Conversely, the decoding involves a combination of upsampling, convolutions, and concatenation of skip connections. As a result, the model outputs the flows $\phi_{0 \rightarrow 1}$ and $\phi_{1 \rightarrow 0}$, each warping $I_0$ to resemble $I_1$ and $I_1$ to resemble $I_0$, respectively.

\subsection{Reconstruction model}
\cref{fig:architecture recon} describes the architecture of the reconstruction model, based on 3D-UNet~\cite{ronneberger2015u}. We employ a single image \(\hat I^{cand}_{0 \rightarrow t} \oplus \hat I^{cand}_{1 \rightarrow t}\), which is a weighted sum of two candidate images, in conjunction with three levels of multi-resolution features, each possessing channel dimensions of 4, 8, and 16, respectively. The model's first encoder layer receives an input composed of two channel-wise concatenated warped features and an image \(\hat I^{cand}_{0 \rightarrow t} \oplus \hat I^{cand}_{1 \rightarrow t}\). Advancing to the subsequent layers, the model concatenates features of half and quarter resolutions at the second and third encoder layers. Thereafter, the model returns the image difference $\Delta \hat I$, which will be added to the input to acquire the final estimated image $\hat I_{t}$. The architecture of the reconstruction model follows details similar to those of the flow calculation model. 

\subsection{Additional training details}
In our training process, we employ the Adam optimizer~\cite{kingma2014adam} with a learning rate \(2 \times 10^{-4}\) for 200 epochs, configuring the batch size as 1. For instance-specific optimization, models are fine-tuned for 100 epochs on the given test sample while maintaining the same experimental settings as in the previous training. The results are presented in a straightforward setup, with all loss coefficients uniformly set to 1.

\newpage
\begin{figure*}
  \centering
  \includegraphics[width=\textwidth]{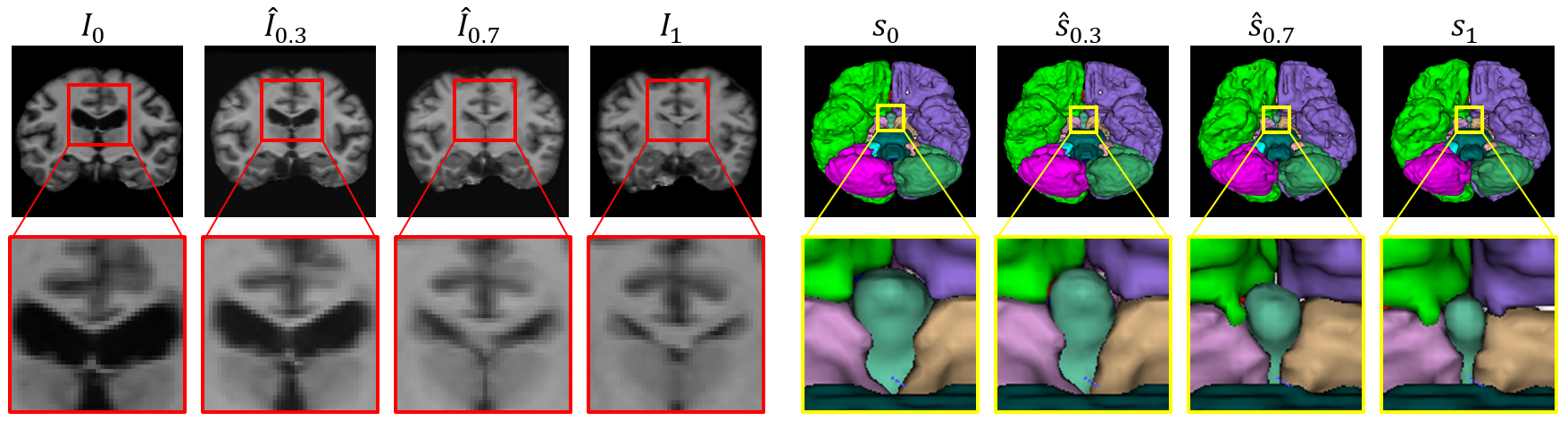}
  \vspace{-0.5cm}
  \caption{Visualization of data augmentation using our approach. Given \(I_0, I_1\) and \(s_0, s_1\), we report the generated image and label when \(t=0.3, 0.7\). This is a visualization based on data from the OASIS dataset.}
  \vspace{-0.3cm}
  \label{fig:augmentation vis}
\end{figure*}


\section{Downstream Task}
\label{sec: augmentation}

\subsection{Method}
We propose an effective 3D data augmentation technique based on our interpolation framework. To extend the interpolation task to 3D data augmentation, we generate new data by inputting randomly selected pairs of 3D images from the training dataset that share common types of segmentation labels. Here, we utilize time \(t\) as an interpolation degree for augmentation. Furthermore, inspired by previous works~\cite{voxelmorph2019, chen2022transmorph}, we incorporate the segmentation labels as supplementary information to enrich the augmented dataset.

Let \(s_0\) and \(s_1\) represent the organ segmentation of \(I_0\) and \(I_1\). When calculating flow fields, we only use \(I_0\) and \(I_1\), excluding segmentation labels. Using the calculated flows, we calculate \(\hat s_{t_1 \rightarrow 0}^{cand}, \; \hat s_{t_2 \rightarrow 0}^{cand}, \; \hat s_{t_2 \rightarrow 1}^{cand}\) and \(\hat s_{t_3 \rightarrow 1}^{cand}\) similar to the procedure of image. Finally, we ensure that \(\hat s_{t_1 \rightarrow 0}^{cand}\) and \(s_{t_2 \rightarrow 0}^{cand}\) have cycle consistency between \(s_0\), while \(s_{t_2 \rightarrow 1}^{cand}\) and \(s_{t_2 \rightarrow 1}^{cand}\) have cycle consistency with \(s_1\).

When labels are used during training, we expand the segmentation map into \(K\) binary masks to enable backpropagation, where \(K\) represents the total number of labels in the segmentation maps. Since Dice score~\citep{dice1945measures} is commonly used to quantify optical flow performance~\cite{voxelmorph2019, chen2022transmorph}, we directly minimized the
Dice loss~\citep{milletari2016v}.

\subsection{Experimental setting}

\paragraph{Datasets.}
For the segmentation dataset for augmentation, three 3D medical datasets are used. OASIS~\cite{hering2022learn2reg} is a brain dataset comprising 414 T1-weighted MRI scans and the corresponding segmentation labels for 36 organs, including the background label released from VoxelMorph~\cite{voxelmorph2019}. IXI~\footnote{https://brain-development.org/ixi-dataset/} is another brain MRI dataset with segmentation labels for 31 organs, including the background ~\cite{chen2022transmorph} released from TransMorph~\cite{chen2022transmorph}. All the brain MRI scans are skull-stripped and resized to $128 \times 128 \times 128$. In both datasets, the first 20 samples are used for training, while the rest are included in the test set. Lastly, MSD-Heart~\cite{simpson2019large} is an MRI dataset with one label (excluding background) and resized to $128 \times 128 \times 64$. Since MSD-Heart has only 20 data, we use 10 data for training and 10 for testing with background loss.

\input{table/segmentation}

\paragraph{Segmentation models.}
To perform 3D segmentation, we utilize three publicly available models from MONAI package\footnote{https://monai.io/}: 3D-UNet~\cite{ronneberger2015u}, VNet~\cite{abdollahi2020vnet}, and UNETR~\cite{hatamizadeh2022unetr}. The segmentation models are trained for 15,000 iteration steps the final Dice score at the last iteration is recorded. Adam optimizer~\cite{kingma2014adam} with an initial learning rate $1 \times 10^{-4}$ is used, and batch size is set to 1. For loss function, the weighted sum of Dice~\cite{milletari2016v} and Cross Entropy~\cite{shannon1948mathematical} losses is used. For augmented data generation, which expands the original dataset size by a factor of ten, we employed alpha sampling ratios of $t = 0.1, 0.2, \ldots, 1.0$.

\subsection{Result}
We have successfully generated pairs of images and labels, as illustrated in~\cref{fig:augmentation vis}. Detailed results presented in~\cref{tab:segmentation result} reveal that our approach consistently outperforms competing methods, delivering superior performance across a diverse range of conditions. This includes variations in dataset types and the use of different segmentation models, underscoring the robustness and versatility of our methodology.

\newpage
\begin{table*}[t]
  \centering
  \begin{adjustbox}{width=0.65\textwidth}
  \footnotesize
  \begin{tabular}{lcccccccc}
    \toprule
    \multirow{2}{*}{\textbf{Dataset}} & \multicolumn{3}{c}{\textbf{Loss function}} & \multirow{2}{*}{PSNR \(\uparrow\)} & \multirow{2}{*}{ NCC \(\uparrow\)} & \multirow{2}{*}{SSIM \(\uparrow\)} & \multirow{2}{*}{NMSE \(\downarrow\)} & \multirow{2}{*}{ LPIPS \(\downarrow\)} \\
    \cline{2-4}
     & \(\mathcal{L}_{warp}\) & \(\mathcal{L}_{image}\) & \(\mathcal{L}_{reg}\) &  & &  &  &\\
    \midrule
    \multirow{3}{*}{\textbf{Cardiac}} & \ding{51} & \ding{51} & & 33.01 & 0.563 & 0.975 & 2.679 & 1.076 \\
    & \ding{51} & & \ding{51} & 33.16 & 0.562 & 0.975 & 2.691 & 1.194 \\
    \cdashline{2-9}
    & \ding{51} & \ding{51} & \ding{51} & \textbf{33.57} & \textbf{0.565} & \textbf{0.977} & \textbf{2.409} & \textbf{1.134} \\
    \bottomrule
  \end{tabular}
  \end{adjustbox}
  \vspace{-0.3cm}
  \caption{Ablation results of loss terms. \(\mathcal{L}_{image}\) and \(\mathcal{L}_{reg}\) are components of \(\mathcal{L}_{cyc}\). NMSE and LPIPS are written in units of \(10^{-2}\).}
  \label{tab:loss_ablation}
\vspace{-0.15cm}
\end{table*}

\begin{table*}[t]
  \centering
  \begin{adjustbox}{width=0.55\textwidth}
  \footnotesize
  \begin{tabular}{lccccccc}
    \toprule
    \multirow{2}{*}{\textbf{Dataset}} & \multicolumn{2}{c}{{\(\bm{\mathcal{L}_{image}}\)}} & \multirow{2}{*}{PSNR \(\uparrow\)} & \multirow{2}{*}{ NCC \(\uparrow\)} & \multirow{2}{*}{SSIM \(\uparrow\)} & \multirow{2}{*}{NMSE \(\downarrow\)} & \multirow{2}{*}{ LPIPS \(\downarrow\)} \\
    \cline{2-3}
     & \(NCC\) & \(\rho\) &  &  & &  & \\
    \midrule
    \multirow{3}{*}{\textbf{Cardiac}} & \ding{51} & & 33.55 & \textbf{0.565} & \textbf{0.977} & \textbf{2.406} & 1.189 \\
    & & \ding{51} & 33.50 & \textbf{0.565} & \textbf{0.977} & 2.437 & 1.316 \\
    \cdashline{2-8}
    & \ding{51} & \ding{51}  & \textbf{33.57} & \textbf{0.565} & \textbf{0.977} & 2.409 & \textbf{1.134} \\
    \bottomrule
  \end{tabular}
  \end{adjustbox}
  \vspace{-0.3cm}
  \caption{Ablation results of loss terms. NMSE and LPIPS are written in units of \(10^{-2}\). \(\mathcal{L}_{image}\) is used for warping loss and cyclic loss, and \(\rho\) stands for Charbonnier loss.}
  \label{tab:loss_ablation_image}
\vspace{-0.15cm}
\end{table*}

\begin{table*}[t!]
  \centering
  \begin{adjustbox}{width=0.6\textwidth}
  \footnotesize
  \begin{tabular}{lccccccc}
    \toprule
     \textbf{Dataset} & \textbf{Feature extractor} & PSNR \(\uparrow\) & NCC \(\uparrow\) & SSIM \(\uparrow\) & NMSE \(\downarrow\) & LPIPS \(\downarrow\) \\
    \midrule
    \multirow{5}{*}{\textbf{Cardiac}} & None & 33.53 & \textbf{0.565} & \textbf{0.977} & 2.410 & 1.163\\
    & Edge detection & 33.49 & \textbf{0.565} & \textbf{0.977} & 2.434 & \textbf{1.101} \\
    & U-Net & 33.50 & \textbf{0.565} & \textbf{0.977} & 2.445 & 1.151 \\
    & Single-scale CNN & 33.49 & 0.564 & \textbf{0.977} & 2.448 & 1.116\\
    \cdashline{2-7}
    & Multi-scale CNN  & \textbf{33.57} & \textbf{0.565} & \textbf{0.977} & \textbf{2.409} & 1.134 \\
    \bottomrule
  \end{tabular}
  \end{adjustbox}
  \vspace{-0.3cm}
  \caption{The ablation results for the feature extraction module. Extract type ``None" indicates not using feature extraction.}
  \label{tab:feature_ablation}
\vspace{-0.4cm}
\end{table*}

\section{Additional experimental results}
\label{sec: supple additional experiment}

We further substantiate our methodology through a series of ablation studies designed to broaden the empirical results. All reported outcomes represent the values derived from three distinct experimental runs.

\subsection{Ablation studies of loss term}
The ablation results of loss terms conducted on the ACDC dataset are summarized in \cref{tab:loss_ablation} and \cref{tab:loss_ablation_image}. As indicated in \cref{tab:loss_ablation}, integrating each component of cyclic loss, which are $\mathcal{L}_{image}$ and $\mathcal{L}_{reg}$, significantly improves the performance of intermediate image synthesis. Furthermore, \cref{tab:loss_ablation_image} demonstrates that the combined application of NCC and Charbonnier losses leads to a performance improvement compared to the application of each loss term independently.

\subsection{Ablation studies of feature extractor model}
The \cref{tab:feature_ablation} presents the results of ablation studies on the feature extraction model, conducted on the ACDC dataset. In our comparative analysis, we demonstrate that our feature extraction methodology exhibits superior performance compared to scenarios where no feature extraction model is implemented. Additionally, we explored alternative methods of feature extraction, including: (1) using the Canny edge detector, (2) employing a simple U-Net architecture, and (3) utilizing a CNN module with single-scale warped images. Our approach outperformed other feature extraction modules in overall metric aspects. Moreover, some metrics in those modules showed performance worse than cases where no feature extraction was applied.

\subsection{Additional qualitative results}
\label{subsec: supple qualitative}
\begin{figure*}[h]
  \centering
  \includegraphics[width=.96\textwidth]{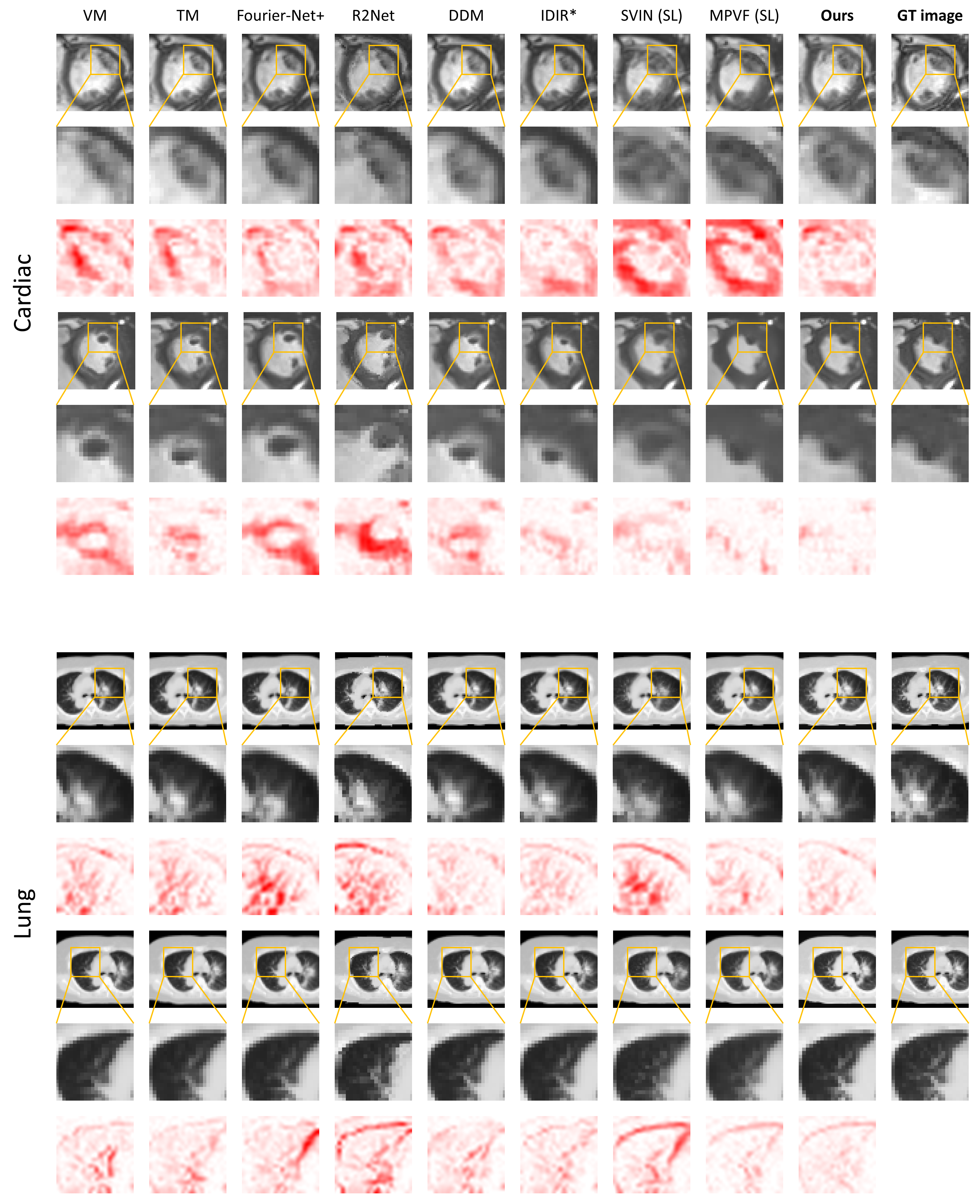}
  \caption{Additional visualization examples demonstrating our proposed method's effectiveness for 4D interpolation. The model marked with an `*' is trained exclusively on the test set, while models marked with `(SL)' are trained using supervised learning. Every third row shows the difference between each model and the ground truth image, where greater pixel value indicates a larger divergence from the ground truth.}
  \label{fig:comparison3}
\end{figure*}

We present a series of additional qualitative results in \cref{fig:comparison3}. Our approach demonstrate the superior results against various baseline methods. This not only underscores our method's enhanced alignment and coordination but also showcases its ability to generate outcomes that are more accurate and realistic. The visual evidence presented here plays a crucial role in substantiating the quantitative metrics we have reported, offering a holistic view of our model's capabilities in real-world scenarios.

\subsection{Visualization for extrapolation}
\begin{figure*}[h]
  \centering
  \includegraphics[width=.95\textwidth]{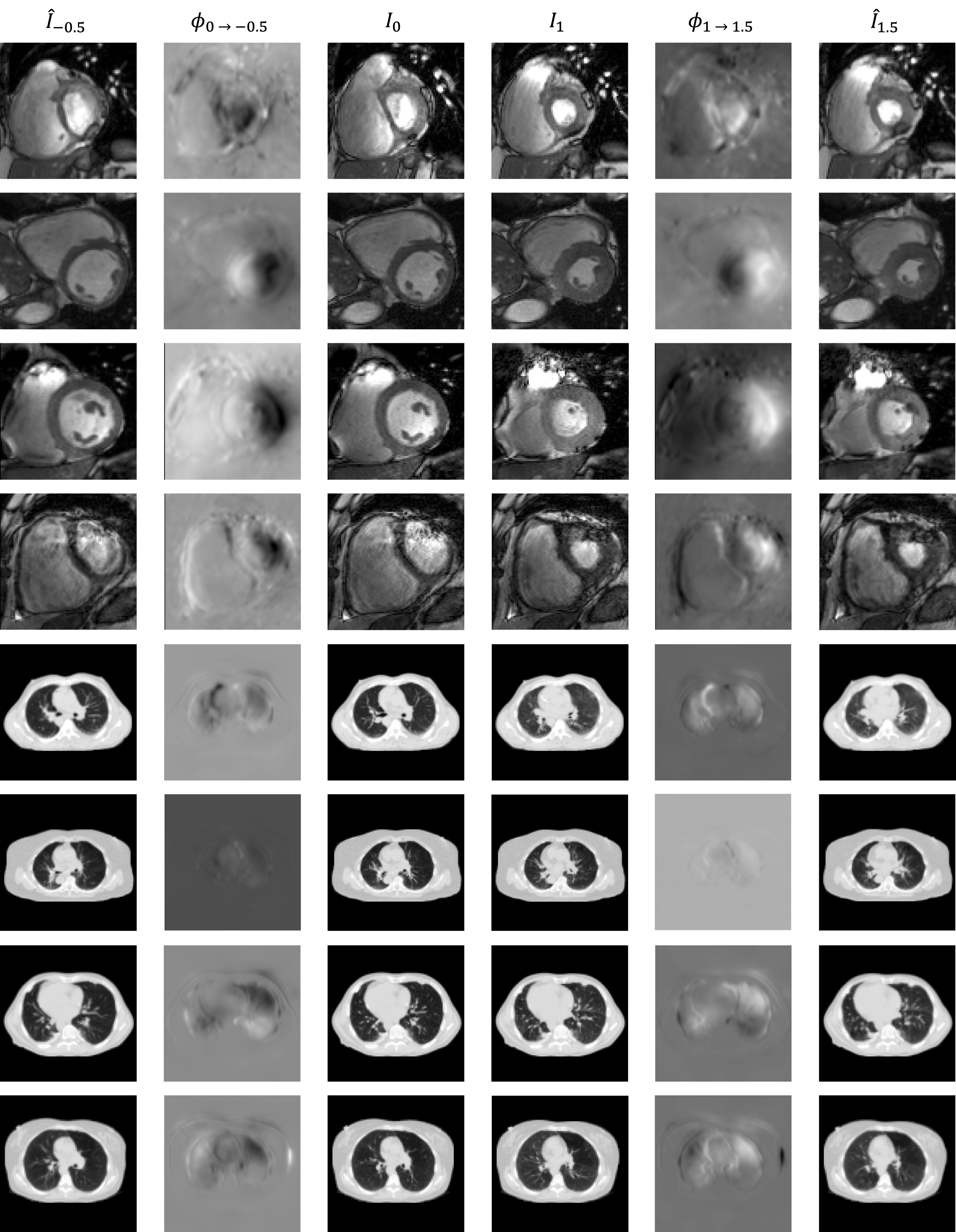}
  \caption{Extrapolation examples for the cardiac and lung datasets. The optical flows presented below pertain to the x-axis direction.}
  \label{fig:extrapolation}
\end{figure*}

The \cref{fig:extrapolation} visualizes the extrapolation results, particularly for $\hat I_{-0.5}$ and $\hat I_{1.5}$, along with the corresponding optical flow and source images $I_0$ and $I_1$. These images represent the most extreme cases of extrapolation in our study. To ensure the credibility and real-world applicability of the results, they have been rigorously examined by a board-certified radiation oncologist. The evaluation focused on determining whether the extrapolated images exhibit any excessive or unnatural changes that could undermine their practical utility. This ensures that using extrapolation in our method does not present significant complications.

\subsection{Visualization results for sequential 4D images}
\begin{figure*}[h]
     \centering
     \begin{subfigure}[h]{\textwidth}
         \centering
         \includegraphics[width=.75\textwidth]{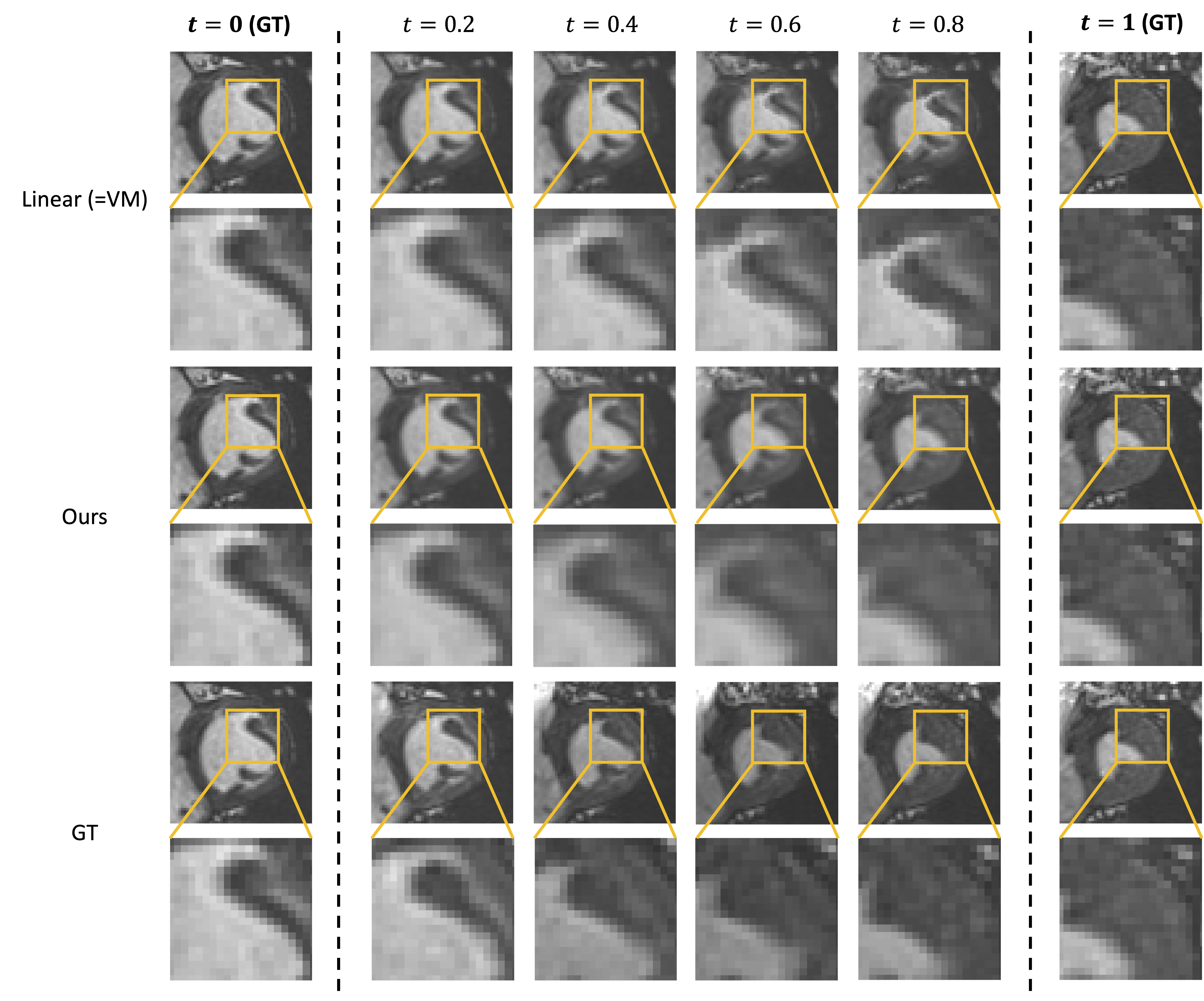}
         \caption{Example of the cardiac dataset.}
         \label{fig:series_cardiac}
     \end{subfigure} \\
     \begin{subfigure}[h]{\textwidth}
         \centering
         \includegraphics[width=.75\textwidth]{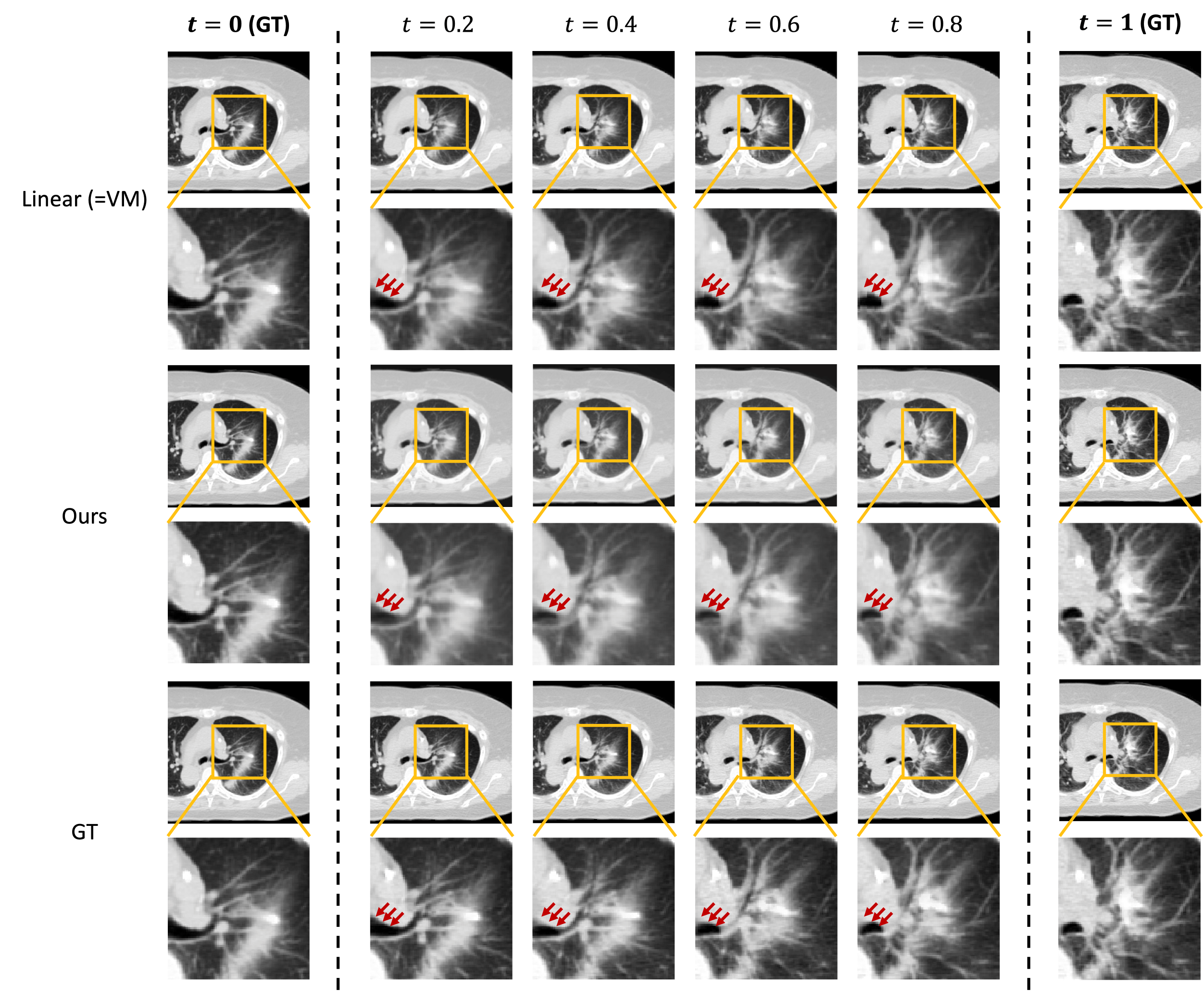}
         \caption{Example of the lung dataset.}
         \label{fig:series_lung}
     \end{subfigure} \\
    \caption{Qualitative results on the prediction of 4D image series over time. For lung images, we present the results in high resolution by upsampling the size of the registration field by a factor of four. In the provided figure, the first and last columns represent the ground truth images. Our model demonstrates a superior ability to capture the fine-grained structures like left main bronchus (indicated by red arrows) compared to the baseline.}
    \label{fig:qual_serial}
\end{figure*}

\cref{fig:qual_serial} visualizes the prediction results over time for the entire 4D sequence. As the baseline results, we introduce the interpolated images through the application of linear scaling to VoxelMorph, which serves as the backbone registration model within our framework. It can be observed that our approach more effectively captures fine-grained details and predicts the ground truth compared to the baseline.

\clearpage

\clearpage

\clearpage
\newpage 

%% file: table/segmentation.tex
\begin{table}[t]
  \centering
  \begin{adjustbox}{width=.8\linewidth}
  \begin{tabular}{lccc}
    \toprule
    \textbf{Method} & OASIS & IXI & MSD-Heart\\
    \midrule
    Vanilla & 0.821 & 0.801 & 0.755 \\
    \hdashline
    VM~\cite{voxelmorph2019} & 0.825 & 0.813 & 0.803 \\
    TM~\cite{chen2022transmorph} & 0.831 & 0.810 & 0.773 \\
     Fourier-Net+~\cite{jia2023fourier}& 0.822 & 0.802 & 0.809 \\
     R2Net~\cite{joshi2023r2net}& 0.621 & 0.688 & 0.789 \\
    DDM~\cite{kim2022diffusion} & 0.826 & 0.806 & 0.818 \\
    \textbf{Ours (w/o inst opt.)} & \textbf{0.843} & \textbf{0.818} & \textbf{0.831} \\

    \bottomrule
  \end{tabular}
  \end{adjustbox}
  \vspace{-0.3cm}
  \caption{Segmentation results on three datasets. Experiments are conducted by adding augmentation data at a scale of 10x to the real data. Dice score is used as the averaged performance metric for three segmentation models.}
  \vspace{-0.3cm}
  \label{tab:segmentation result}
\end{table}